\documentclass[a4paper, 10pt, twocolumn]{article}

\usepackage[utf8]{inputenc}

\makeatletter
\newcommand{\LoadPackagesNow}{}
\newcommand{\LoadPackageLater}[2][]{%
   \g@addto@macro{\LoadPackagesNow}{%
      \usepackage[#1]{#2}%
   }%
}
\makeatother

\usepackage{xspace}
\usepackage{xargs}
\usepackage{ifthen}
\usepackage{etoolbox}

\usepackage{calc}
\usepackage[top=1.5cm, left=1.5cm, right=1.5cm, bottom=1.5cm]{geometry}
\usepackage{framed}
\usepackage{mdframed}
\usepackage{pbox}
\usepackage{enumitem}
\usepackage{array}
\usepackage{multirow}
\usepackage{longtable}
\usepackage{hhline}
\usepackage{afterpage}
\usepackage{pdflscape}
\usepackage{rotating}
\usepackage{booktabs}
\usepackage{setspace}
\usepackage[
   bottom,      % Footnotes appear always on bottom. This is necessary
   stable,      % Make footnotes stable in section titles
   multiple,    % rearrange multiple footnotes intelligent in the text.
   flushmargin,
   hang,
]{footmisc}
\usepackage{relsize}
\usepackage[normalem]{ulem}      
\usepackage{soul}		           
\usepackage{url}
\usepackage{lipsum}
\usepackage[table]{xcolor}
\usepackage{tikz}
\usetikzlibrary{arrows,shapes,calc,decorations.pathreplacing,intersections,quotes,angles,positioning,fit,petri,through,plotmarks}
\tikzstyle{blackdot}=[shape=circle,fill=black,minimum size=1mm,inner sep=0pt,outer sep=0pt]
\usepackage[%
]{graphicx}
\usepackage{epstopdf}
\usepackage{wrapfig}
\usepackage{caption}
\usepackage{subcaption}
\usepackage[
   tbtags,    % 'Top-or-bottom tags': For a split equation, place equation numbers level
   sumlimits,  %(default) Place the subscripts and superscripts of summation
   nointlimits, % (default) Opposite of intlimits.
   namelimits, % (default) Like sumlimits, but for certain 'operator names' such as
   reqno,     % Place equation numbers on the right.
]{amsmath} %

\usepackage{amsfonts}
\usepackage{mathrsfs} %% Schreibschriftbuchstaben für den Mathematiksatz (nur Großbuchstaben)
\usepackage{dsfont}
\usepackage{amssymb}
\usepackage{units}
\usepackage{stmaryrd}
\LoadPackageLater{amsthm}
\LoadPackageLater{thmtools}
\usepackage[fixamsmath,disallowspaces]{mathtools}
\mathtoolsset{showonlyrefs}
\mathtoolsset{centercolon=true}
\usepackage{bm} 
\makeatletter
\g@addto@macro\bfseries{\boldmath}
\makeatother
\allowdisplaybreaks[4]
\numberwithin{equation}{section}
\usepackage{arydshln}
\usepackage{diagbox}
\usepackage{makecell}

\usepackage[toc]{appendix}

\usepackage{csquotes}
\usepackage[
	bibstyle=alphabetic,
	citestyle=alphabetic,
	sorting=ynt,
	sortcites=true,
	giveninits=true,
	maxbibnames=99,
	backend=biber,
	maxalphanames=5,
]{biblatex}

\renewbibmacro{in:}{}
\DeclareFieldFormat{pages}{#1}
\usepackage[textsize=tiny,english,colorinlistoftodos,
	disable,
	]{todonotes}
\definecolor{pdfurlcolor}{rgb}{0,0,0.6}
\definecolor{pdffilecolor}{rgb}{0.7,0,0}
\definecolor{pdflinkcolor}{rgb}{0,0,0.6}
\definecolor{pdfcitecolor}{rgb}{0,0,0.6}
\usepackage[
  colorlinks=true,         % Links erhalten Farben statt Kaeten
  urlcolor=pdfurlcolor,    % \href{...}{...} external (URL)
  filecolor=pdffilecolor,  % \href{...} local file
  linkcolor=pdflinkcolor,  %\ref{...} and \pageref{...}
  citecolor=pdfcitecolor,  %
  raiselinks=true,			 % calculate real height of the link
  breaklinks,              % Links berstehen Zeilenumbruch
  verbose,
  hyperindex=true,         % backlinkex index
  linktocpage=true,        % Inhaltsverzeichnis verlinkt Seiten
  hyperfootnotes=false,     % Keine Links auf Fussnoten
  bookmarks=true,          % Erzeugung von Bookmarks fuer PDF-Viewer
  bookmarksopenlevel=2,    % Gliederungstiefe der Bookmarks
  bookmarksopen=true,      % Expandierte Untermenues in Bookmarks
  bookmarksnumbered=true,  % Nummerierung der Bookmarks
  bookmarkstype=toc,       % Art der Verzeichnisses
  plainpages=false,        % Anchors even on plain pages ?
  pageanchor=true,         % Pages are linkable
  pdfdisplaydoctitle=true, % Dokumententitel statt Dateiname im Fenstertitel
  pdfstartview=FitH,       % Dokument wird Fit Width geaefnet
  pdfpagemode=UseOutlines, % Bookmarks im Viewer anzeigen
  pdfpagelabels=true,           % set PDF page labels
  pdfpagelayout=OneColumn, % zweiseitige Darstellung: ungerade Seiten
]{hyperref}

\LoadPackagesNow

\newcommand{\ifargdef}[3][{}]{\ifthenelse{\equal{#2}{}}{#1}{#3}}

\newlength{\hangind}
\newcommand{\myhangindent}[1]{\settowidth{\hangind}{\widthof{#1}}\hangindent=\the\hangind}

\usepackage{amsfonts,amssymb,amsmath,amsthm}
\usepackage{graphicx}

\newenvironment{highlight}{\begin{quote}\itshape}{\end{quote}}

\newtheoremstyle{claim}
	{\topsep}{\topsep}%
	{\itshape}%         Body font
	{}%         Indent amount (empty = no indent, \parindent = para indent)
	{}% Thm head font
	{}%        Punctuation after thm head
	{.5em}%     Space after thm head (\newline = linebreak)
	{{\bfseries\boldmath\thmname{#1} \thmnumber{#2}} \thmnote{(#3)}}%         Thm head spec

\newtheoremstyle{definition}
	{\topsep}{\topsep}%
	{}%         Body font
	{}%         Indent amount (empty = no indent, \parindent = para indent)
	{}% Thm head font
	{}%        Punctuation after thm head
	{.5em}%     Space after thm head (\newline = linebreak)
	{\textbf{\thmname{#1} \thmnumber{#2}} \thmnote{(#3)}}%         Thm head spec
	
\newtheoremstyle{algorithm}
	{\topsep}{\topsep}%
	{}%         Body font
	{}%         Indent amount (empty = no indent, \parindent = para indent)
	{\bfseries\boldmath}% Thm head font
	{}%        Punctuation after thm head
	{\newline}%     Space after thm head (\newline = linebreak)
	{\thmname{#1} \thmnumber{#2} \thmnote{(#3)}}%         Thm head spec

\mdfdefinestyle{boxed}{innertopmargin =6pt, splittopskip = \topskip, skipbelow=6pt, skipabove=12pt}

\mdfdefinestyle{emphframe}{linecolor=black, innertopmargin = 3pt, splittopskip = \topskip, skipbelow=6pt, skipabove=6pt}

\declaretheorem[style=claim,numberwithin=section]{theorem}

\declaretheorem[style=definition,sibling=theorem]{definition}

\declaretheorem[style=algorithm,sibling=theorem,%
	preheadhook={\begin{mdframed}[style=emphframe] \setcounter{mpfootnote}{\value{footnote}}},%
	postfoothook=\end{mdframed}]{experiment}
\declaretheorem[style=algorithm,sibling=theorem,%
	preheadhook={\begin{mdframed}[style=emphframe] \setcounter{mpfootnote}{\value{footnote}}},%
	postfoothook=\end{mdframed}]{algorithm}

\newcommand{\opleft}[1]{\mathopen{}\left#1}
\newcommand{\opright}[1]{\right#1\mathclose{}}
\newcommandx{\braces}[4]{%
\ifstrequal{#3}{normal}{#1#4#2}{%
\ifstrequal{#3}{auto}{\left#1#4\right#2}{%
\ifstrequal{#3}{opauto}{\opleft#1#4\opright#2}{%
#3#1#4#3#2}}}%
}
\newcommandx{\opannot}[3][3=\downarrow]{\stackrel{\mathclap{\substack{#1 \\ #3 \vspace{2pt}}}}{#2}}
\newcommandx{\lineannot}[3][3=\rightarrow]{\mathllap{\boxed{\text{\textsmaller{#1}}} #3} #2}
\newcommandx{\multilineannot}[4][4=\rightarrow]{\mathllap{\boxed{\parbox{#1}{\RaggedRight\textsmaller{#2}}} #4} #3}
\newcommand{\R}{\mathbb{R}} % real numbers
\newcommand{\suchthat}[1][normal]{\ifstrequal{#1}{normal}{\mid}{#1|}} % seperator in sets (#1op = size)

\newcommand{\cardinality}[1]{\abs{#1}} % cardinality of a set
\newcommandx{\intvcl}[3][1=normal]{\braces{[}{]}{#1}{#2, #3}} % closed interval (#1op=size, #2=left bound, #3=right bound)
\newcommandx{\intvop}[3][1=normal]{\braces{(}{)}{#1}{#2, #3}} % open interval
\newcommandx{\intvclop}[3][1=normal]{\braces{[}{)}{#1}{#2, #3}} % half-open interval (right)
\newcommandx{\intvopcl}[3][1=normal]{\braces{(}{]}{#1}{#2, #3}} % half-open interval (left)
\newcommandx{\abs}[2][1=normal]{\braces{\lvert}{\rvert}{#1}{#2}} % absolute value
\newcommandx{\ceil}[2][1=normal]{\braces{\lceil}{\rceil}{#1}{#2}} % ceil
\newcommandx{\floor}[2][1=normal]{\braces{\lfloor}{\rfloor}{#1}{#2}} % floor
\newcommandx{\round}[2][1=normal]{\braces{\llbracket}{\rrbracket}{#1}{#2}} % round
\newcommandx{\der}[1]{D^{#1}} % differential operator (#1 = multiindex)
\newcommandx{\gradient}{\nabla} % gradient
\newcommandx{\partder}[4][1={},4={}]{\frac{\partial^{#4} #2}{\partial #3^{#4}}\ifargdef{#1}{\Big|_{#1}}} % partial derivative (#1=point of evaluation, #2=function, #3=variable, #4=order)
\newcommandx{\integ}[4][1={},2={}]{\int_{#1}^{#2} #3 \, #4} % integral (#1op=lower bound, #2op=upper bound, #3=integrand, #4=differential form)
\newcommandx{\asympffaster}[2][1=normal]{o\braces{(}{)}{#1}{#2}} % asymptotically faster (proper) (#1op=size)
\newcommandx{\asympfaster}[2][1=normal]{O\braces{(}{)}{#1}{#2}} % asymptotically faster
\newcommandx{\asympeq}[2][1=normal]{\Theta\braces{(}{)}{#1}{#2}} % asymptotically equal
\newcommandx{\asympsslower}[2][1=normal]{\omega\braces{(}{)}{#1}{#2}} % asymptotically slower (proper)
\newcommandx{\asympslower}[2][1=normal]{\Omega\braces{(}{)}{#1}{#2}} % asymptotically slower

\newcommandx{\norm}[2][1=normal]{\braces{\|}{\|}{#1}{#2}} % norm
\renewcommandx{\sp}[3][1=normal]{\braces{\langle}{\rangle}{#1}{#2, #3}} % inner product (#1op=size, #2=left, #3=right)
\newcommandx{\End}[2][2={}]{\mathcal{L}\opleft( #1 \ifargdef{#2}{, #2} \opright)} % endomorphism (#1=from, #2op=to)
\renewcommand{\vec}[1]{\boldsymbol{#1}} % vectors in boldface

\newcommandx{\measure}[2][1=normal]{\operatorname{vol}\braces{(}{)}{#1}{#2}} % Lebesgue-measure/volume of a set
\DeclareMathOperator{\supp}{supp} % support
\newcommandx{\Leb}[3][1={},3=normal]{L^{#2}\ifargdef{#1}{\braces{(}{)}{#3}{#1}}{}} % Lebesgue spaces (#1op=set, #2=exponent)
\newcommandx{\Lebnorm}[4][1=normal,3={2},4={}]{\norm[#1]{#2}_{#3}} % Lebesgue norm (#1op=size, #2=content, #3op=exponent, #4op=set)
\renewcommandx{\l}[3][1={},3=normal]{\ell^{#2}\ifargdef{#1}{\braces{(}{)}{#3}{#1}}} % lp sequence spaces (#1op=set, #2=exponent)
\newcommandx{\lnorm}[4][1=normal,3={2},4={}]{\norm[#1]{#2}_{#3}} % lp norm (#1op=size, #2=content, #3op=exponent, #4op=set)
\newcommandx{\Smooth}[4][1={},3={},4=normal]{C_{#3}^{#2}\ifargdef{#1}{\braces{(}{)}{#4}{#1}}} % space of differentiable functions (#1op=set, #2=order, #3op=modifier)
\newcommandx{\Schwartz}[2][1={},2=normal]{\mathscr{S}\ifargdef{#1}{\braces{(}{)}{#2}{#1}}} % space of Schwartz functions
\newcommandx{\Schwartzpoly}[2][1=normal]{\braces{\langle}{\rangle}{#1}{\abs[#1]{#2}} } % Schwartz polynomial
\newcommandx{\Tempdistr}[2][1={},2=normal]{\mathscr{S}'\ifargdef{#1}{\braces{(}{)}{#2}{#1}}} % tempered distributions
\newcommandx{\distrinp}[3][1=normal]{\braces{\langle}{\rangle}{#1}{#2, #3}} % evaluation of a tempered distribution (#1op=size, #2=distribution, #3=Schwartz function)
\newcommandx{\ft}[3][1=default,2=auto]{
\ifstrequal{#1}{default}{\widehat{#3}}{
\ifstrequal{#1}{long}{{\braces{(}{)}{#2}{#3}}^{\wedge}}{}}} % Fourier transform (hat-notation) (#1op=long expression mode, #2op=size, #3=content)
\newcommandx{\ift}[3][1=default,2=auto]{
\ifstrequal{#1}{default}{\check{#3}}{
\ifstrequal{#1}{long}{{\braces{(}{)}{#2}{#3}}^{\vee}}{}}} % inverse Fourier transform (hat-notation) (#1op=long expression mode, #2op=size, #3=content)
\DeclareMathOperator{\PolyLog}{PolyLog} % Polylog

\newcommand{\y}{\vec{y}} % measurements
\newcommand{\A}{\vec{A}} % meausrement matrix
\newcommand{\x}{\vec{x}} % vector
\newcommand{\grtr}{\vec{x}^\ast} % groundtruth signal
\newcommandx{\solu}[1][1={}]{\ifargdef[\hat{\vec{x}}]{#1}{\hat{#1}}} % minimizer
\newcommand{\contgrtr}{\mathcal{X}} % continuous signal

\newcommand{\sset}{K} % signal set

\newcommand{\probsuccess}{u} %\epsilon

\newcommandx{\prob}[2][1={},2=normal]{\mathbb{P}\ifargdef{#1}{\braces{[}{]}{#2}{#1}}}

\newcommandx{\mean}[2][1={},2=normal]{\mathbb{E}\ifargdef{#1}{\braces{[}{]}{#2}{#1}}}
\newcommandx{\var}[2][1={},2=normal]{\mathbb{V}\ifargdef{#1}{\braces{[}{]}{#2}{#1}}}
\newcommandx{\Normdistr}[3][1=normal]{\mathcal{N}\braces{(}{)}{#1}{#2, #3}} % Normal distribution
\newcommandx{\normsubg}[2][1=normal]{\norm[#1]{#2}_{\psi_2}} % sub-Gaussian
\newcommandx{\anorm}[3][1=normal,3={\sset}]{\norm[#1]{#2}_{#3}} % atomic norm 
\newcommandx{\pospart}[2][1=auto]{\braces{[}{]}{#1}{#2}_+}

\newcommandx{\ball}[2][1={},2={}]{B_{#1}^{#2}} % lp unit ball
\newcommand{\effdim}[2][{}]{w_{#1}^2(#2)} % effective dimension
\newcommand{\conic}{\wedge} % indicator for conic version
\newcommand{\descset}[1]{\mathcal{D}(#1)} % descent set
\newcommandx{\subdiff}[2][2={}]{\partial #1\ifargdef{#2}{(#2)}{}} % subdifferential
\newcommandx{\clip}[3][1=normal]{\operatorname{clip}\braces{(}{)}{#1}{#2; #3}} % clip function

\newcommand{\s}{s}

\newcommand{\SC}{\Delta}

\newcommand{\jump}{\nu}

\newcommand{\TV}{\nabla} % TV operator

\title{Compressed Sensing with 1D Total Variation: Breaking Sample Complexity Barriers via Non-Uniform Recovery}

\author{Martin Genzel$^1$, Maximilian März$^1$, and Robert Seidel$^{2}$\\
\footnotesize $^1$Technische Universität Berlin, Department of Mathematics, Berlin, Germany.\\ 
\footnotesize  $^2$Technische Universität Berlin, Institute of Software Engineering and Theoretical Computer Science, Berlin, Germany.\
}
\date{\empty} % no need for a date

\renewenvironment{abstract}{\bf\small {\em\ Abstract---}}{}

\begin{document}

\maketitle

\begin{abstract} 
This paper investigates total variation minimization in one spatial dimension for the recovery of gradient-sparse signals from undersampled Gaussian measurements. Recently established bounds for the required sampling rate state that uniform recovery of all $s$-gradient-sparse signals in $\R^n$ is only possible with $m \gtrsim \sqrt{\s n} \cdot \PolyLog(n)$ measurements. 
Such a condition is especially prohibitive for high-dimensional problems, where $s$ is much smaller than $n$. However, previous empirical findings seem to indicate that the latter sampling rate does not reflect the typical behavior of total variation minimization.
Indeed, this work provides a rigorous analysis that breaks the $\sqrt{\s n}$-bottleneck for a large class of natural signals.
The main result shows that non-uniform recovery succeeds with high probability for $m \gtrsim \s \cdot \PolyLog(n)$ measurements if the jump discontinuities of the signal vector are sufficiently well separated.
In particular, this guarantee allows for signals arising from a discretization of piecewise constant functions defined on an interval. 
The present paper serves as a short summary of the main results in our recent work~\cite{Genz2020}.

% The key ingredient of the proof is a novel upper bound for the associated conic Gaussian mean width, which is based on a signal-dependent, non-dyadic Haar wavelet transform. Furthermore, a natural extension to stable and robust recovery is addressed.
\end{abstract}

\section{Introduction}
\label{sec:introduction}

We consider the following inverse problem:  Assume that $\grtr \in \R^n$ denotes a signal of interest in one spatial dimension. It is assumed to be \emph{$\s$-gradient-sparse}, i.e., $\cardinality{\supp( \TV\grtr)} \leq \s$, where $\TV \in \R^{n-1 \times n}$ denotes a \emph{discrete gradient operator}.\footnote{We consider a gradient operator that is based on forward differences and von~Neumann boundary conditions. An extension to other choices is expected to be straightforward.} Instead of having direct access to $\grtr$, the signal is observed via a \emph{linear, non-adaptive measurement process}\footnote{For the sake of simplicity, potential distortions in the measurement process are ignored here, but we emphasize that all results of this work can be made robust against (adversarial) noise.} 
\begin{equation}\label{eq:meas}
	\y = \A \grtr \in \R^m,
\end{equation}
where $\A \in \R^{m \times n}$ is a known \emph{measurement matrix}. 
The methodology of compressed sensing suggests that, under certain conditions, it remains possible to retrieve $\grtr$ from the knowledge of $\y$ and $\A$ even when $m\ll n$. 
Indeed, one of the seminal works of this field by \citeauthor*{candes2006cs}~\cite{candes2006cs} shows that for random Fourier measurements, the recovery of $\grtr$ remains feasible with high probability as long as the number of measurements obeys $m \gtrsim \s \log(n)$, where the `$\gtrsim$'-notation hides a universal constant. 
For the success of this strategy, it is crucial to employ non-linear recovery methods that exploit the a priori knowledge that $\grtr$ is gradient-sparse. 
% There exist numerous greedy methods and convex programs that are designed to accomplish this task efficiently. 
Arguably, the most popular version of 1D total variation (TV) minimization is based on an adaption of the classical basis pursuit, i.e., one solves the convex problem
\begin{equation}\label{eq:intro:tv-1}\tag{$\text{TV-1}$}
	\min_{\x \in \R^n} \lnorm{\TV \x}[1] \quad \text{subject to \quad $\y = \A \x $.}
\end{equation}
The research of the past three decades demonstrates that encouraging a small TV norm often efficaciously reflects the inherent structure of real-world signals. Although not as popular as its counterpart in 2D (e.g., see \cite{rudin_nonlinear_1992,Chambolle1997,Chambolle2004}), TV methods in one spatial dimension find application in many practical scenarios, e.g., see \cite{Little2011,Little2010,Sandbichler2015,Wu2014,Perrone2016}. Furthermore, TV in 1D has frequently been subject of mathematical research \cite{Condat2013,Selesnick2015,Selesnick2012,Mammen1997,Briani2011,Grasmair2007}. 

The main objective of this work is to study the 1D TV minimization problem for the benchmark case of Gaussian random measurements. In a nutshell, we intend to answer the following question:
\begin{highlight}
Assuming that $\A \in \R^{m\times n}$ is a standard Gaussian random matrix, under which conditions is it possible to recover an $\s$-gradient-sparse signal $\grtr \in \R^n$ via TV minimization~\eqref{eq:intro:tv-1} with the near-optimal rate of $m \gtrsim \s\cdot \PolyLog(n)$ measurements?
\end{highlight}

\section{Why Should We Care?}

At first sight, the aforementioned recovery result of \citeauthor*{candes2006cs}~\cite{candes2006cs} seems to deny the relevance of the previous research question. However, we emphasize that their result applies exclusively to random Fourier measurements. Indeed, the TV-Fourier combination allows for a significant simplification of the problem, since the gradient operator is ``compatible'' with the Fourier transform (differentiation is a Fourier multiplier). 

In contrast, the more recent work of \citeauthor{cai_guarantees_2015}~\cite{cai_guarantees_2015} addresses the generic case of Gaussian measurements. However, their main result~\cite[Thm.~2.1]{cai_guarantees_2015} seems to imply a negative answer to the question above: in essence, it shows that the \emph{uniform} recovery of every $s$-gradient-sparse signal by solving~\eqref{eq:intro:tv-1} is possible if and only if the number of measurements obeys
\begin{equation}
 m \gtrsim \sqrt{sn}  \cdot \log (n).
\end{equation}
The conclusion from this result is as surprising as it is discouraging: It suggests that the threshold for successful recovery of $s$-gradient-sparse signals via \eqref{eq:intro:tv-1}  is essentially given by $\sqrt{s n}$-many Gaussian measurements. Remarkably, the latter rate does not resemble the desirable standard criterion ${m \gtrsim \s \cdot \PolyLog (n,\s)}$. 

In Table~\ref{tab:intro}, we have summarized some of the existing guarantees for TV minimization in compressed sensing.  We refer the interested reader to~\cite[Sec.~1.2]{Genz2020} and~\cite{krahmer_total_2017} for a more detailed overview of the relevant literature. 

\begin{table}[ht]
	\renewcommand{\arraystretch}{1.5}
	\begin{center}
		\begin{tabular}{|c||c|c|}
			\hline
			\diagbox[width=1.5cm,height=1cm]{$\A$}{$d$D} &  1D  & $\geq$2D \\
			\hline \hline
			\multirow{2}{*}{\footnotesize Gaussian} & { \footnotesize \textbf{$\s \log^2 (n)$} (non-unif.) \scriptsize [ours]} & \footnotesize $\s \cdot \PolyLog(n,\s)$ \\
			\cdashline{2-2}
			& {\footnotesize $\sqrt{\s n} \cdot \log (n)$ (unif.)  \scriptsize\cite{cai_guarantees_2015}} & \scriptsize\cite{cai_guarantees_2015,Needell2013,Needell2013b} \\
			\hline
			\multirow{2}{*}{\footnotesize Fourier} & \multicolumn{2}{c|}{\footnotesize $\s \cdot \PolyLog(n,\s)$}  \\
			& \multicolumn{2}{c|}{\scriptsize \cite{candes2006cs,Poon2015,Krahmer2014}} \\
			%			\hline
			%			Haar-incoherent & -- & $\s  \log (n^2 /\s)$ {\footnotesize \cite{Needell2013}} & $\s  \log (n^d)$ {\footnotesize\cite{Needell2013b}} \\
			% \hdashline
			\hline
		\end{tabular}
	\end{center}
	\caption{An overview of known asymptotic-order sampling rates for TV minimization in compressed sensing, ignoring universal and model-dependent constants.}
	\label{tab:intro}
\end{table}

\section{Our Contribution}

The main contribution of this work consists in breaking the aforementioned $\sqrt{sn}$-complexity barrier. Taking a non-uniform, signal-dependent perspective, we show that a large class of gradient-sparse signals is already recoverable from $m \gtrsim \s \cdot \PolyLog (n)$ Gaussian measurements.
%\footnote{For the sake of clarity, we will only consider the case of Gaussian measurements in this article. However, our main results can be easily extended to the sub-Gaussian case, i.e., $\A$ has i.i.d.\ isotropic, sub-Gaussian rows; see \cite[Subsec.~6.1]{genzel2017cosparsity} for more details. Apart from that, we emphasize that, despite their practical limitations, (sub-)Gaussian measurement ensembles form a generic and widely accepted benchmark for the analysis of compressed sensing algorithms.} 
Note that such a result does not contradict the findings of \citeauthor{cai_guarantees_2015}~\cite{cai_guarantees_2015}, as these are formulated uniformly across all s-gradient-sparse. Indeed, the $\sqrt{sn}$-rate describes the worst-case performance on the class of all $s$-gradient-sparse signals. We show that a meaningful restriction of this class allows for a significant improvement of the situation, cf.~the numerical experiments of~\cite{cai_guarantees_2015,genzel2017cosparsity}.
With that in mind, our analysis reveals that the separation distance of jump discontinuities of~$\grtr$ is crucial:
\begin{definition}[Separation constant]\label{def:results:msc}
	Let $\grtr \in \R^n$ be a signal with $\s > 0$ \emph{jump discontinuities} such that $\supp(\TV\grtr) = \{\jump_1,\dots,\jump_{\s}\}$ where $0 \eqqcolon \jump_0 < \jump_1 < \dots < \jump_{\s} < \jump_{\s+1} \coloneqq n$.
	We say that $\grtr$ is \emph{$\SC$-separated} for some \emph{separation constant} $\SC > 0$ if
	\begin{equation}\label{eq:results:msc}
	\min_{i \in [\s+1]} \frac{\abs{\jump_i - \jump_{i-1}}}{n} \geq \frac{\SC}{\s+1}.
	\end{equation}
\end{definition}

It is not hard to see that the separation constant can always be chosen such that $(s+1)/n \leq \SC \leq 1$, where larger values of $\SC$ indicate that the gradient support is closer to being equidistant. Indeed, in the (optimal) case of equidistantly distributed singularities, $\SC = 1$ is a valid choice, independently of $s$.  Based on this notion of separation, our main result reads as follows:
\begin{theorem}[Exact recovery via TV minimization]\label{thm:results:exact}
	Let $\grtr \in \R^n$ be a $\SC$-separated signal with $\s>0$ jump discontinuities and $\SC \geq 8 \s / n$.
	Let $\probsuccess > 0$ and assume that $\A \in \R^{m \times n}$ is a standard Gaussian random matrix with
	\begin{equation}\label{eq:results:exact:meas}
	m \gtrsim \SC^{-1} \cdot \s\log^2(n) + \probsuccess^2.
	\end{equation}
	Then with probability at least $1 - e^{-\probsuccess^2/2}$, TV minimization \eqref{eq:intro:tv-1} with input $\y = \A \grtr \in \R^m$ recovers $\grtr$ exactly.
\end{theorem}

% Note that Theorem~\ref{thm:results:exact} provides a \emph{non-uniform} and \emph{signal-dependent} guarantee, in the sense that it concerns the successful recovery of a \emph{fixed} signal $\grtr$ (for a single random draw of $\A$) with a sampling rate that depends on $\grtr$ (in terms of $\SC$). 
% This is in stark contrast to Theorem~\ref{thm:cai}, which addresses recovery of all $s$-gradient-sparse signals simultaneously.

The proof of Theorem~\ref{thm:results:exact} relies on a sophisticated upper bound for the associated conic Gaussian mean width, which is based on a
signal-dependent, non-dyadic Haar wavelet transform. As such, the latter result can be extended to sub-Gaussian measurements as well as stable and robust recovery; see~\cite[Sec.~2.4]{Genz2020} for more details.

\begin{figure}[ht!]
	\centering
	\begin{subfigure}[t]{0.45\linewidth}
		\centering
		\includegraphics[width=\linewidth]{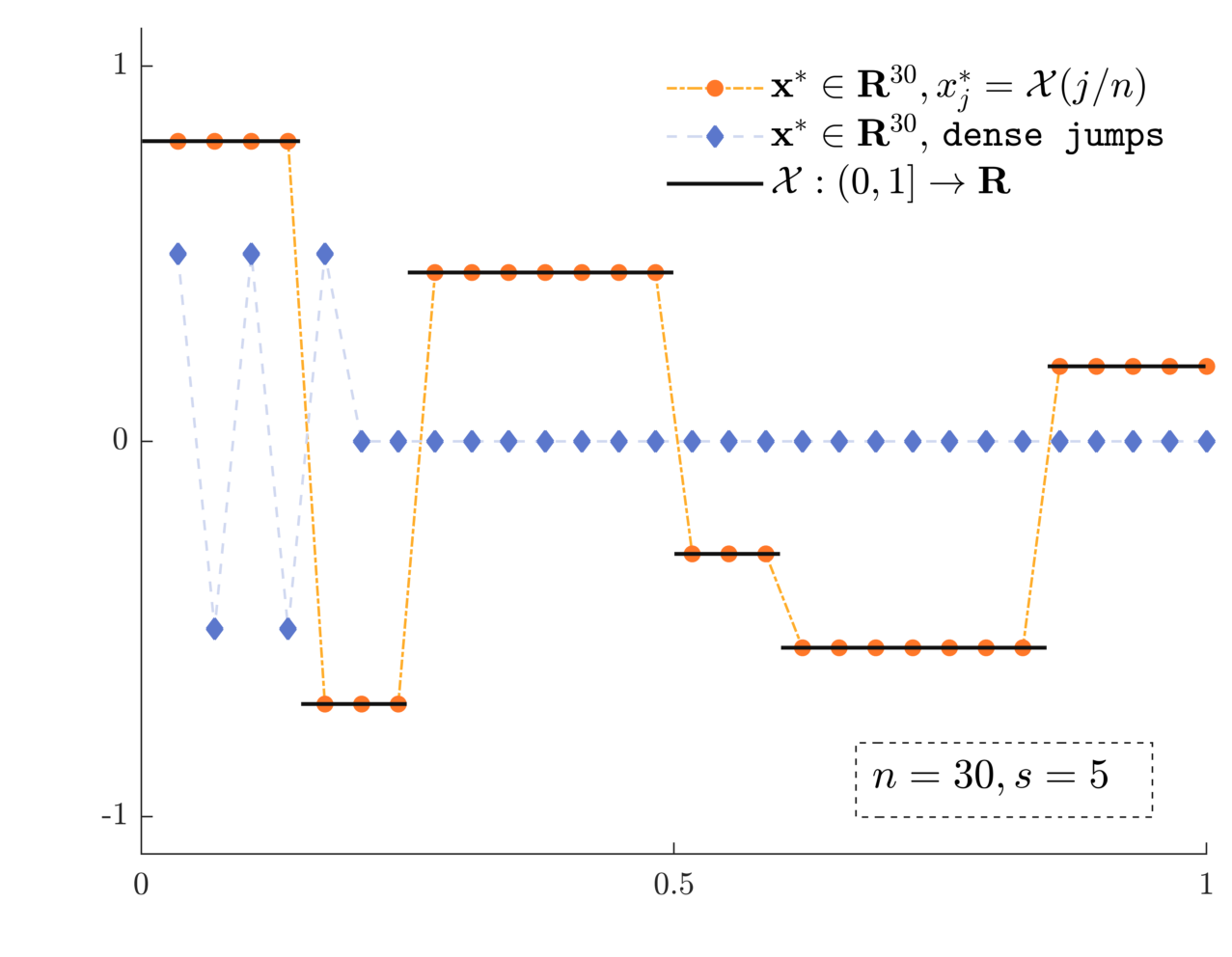}
		\caption{}
		\label{fig:intro:n30}
	\end{subfigure}%
	\qquad
	\begin{subfigure}[t]{0.45\linewidth}
		\centering
		\includegraphics[width=\linewidth]{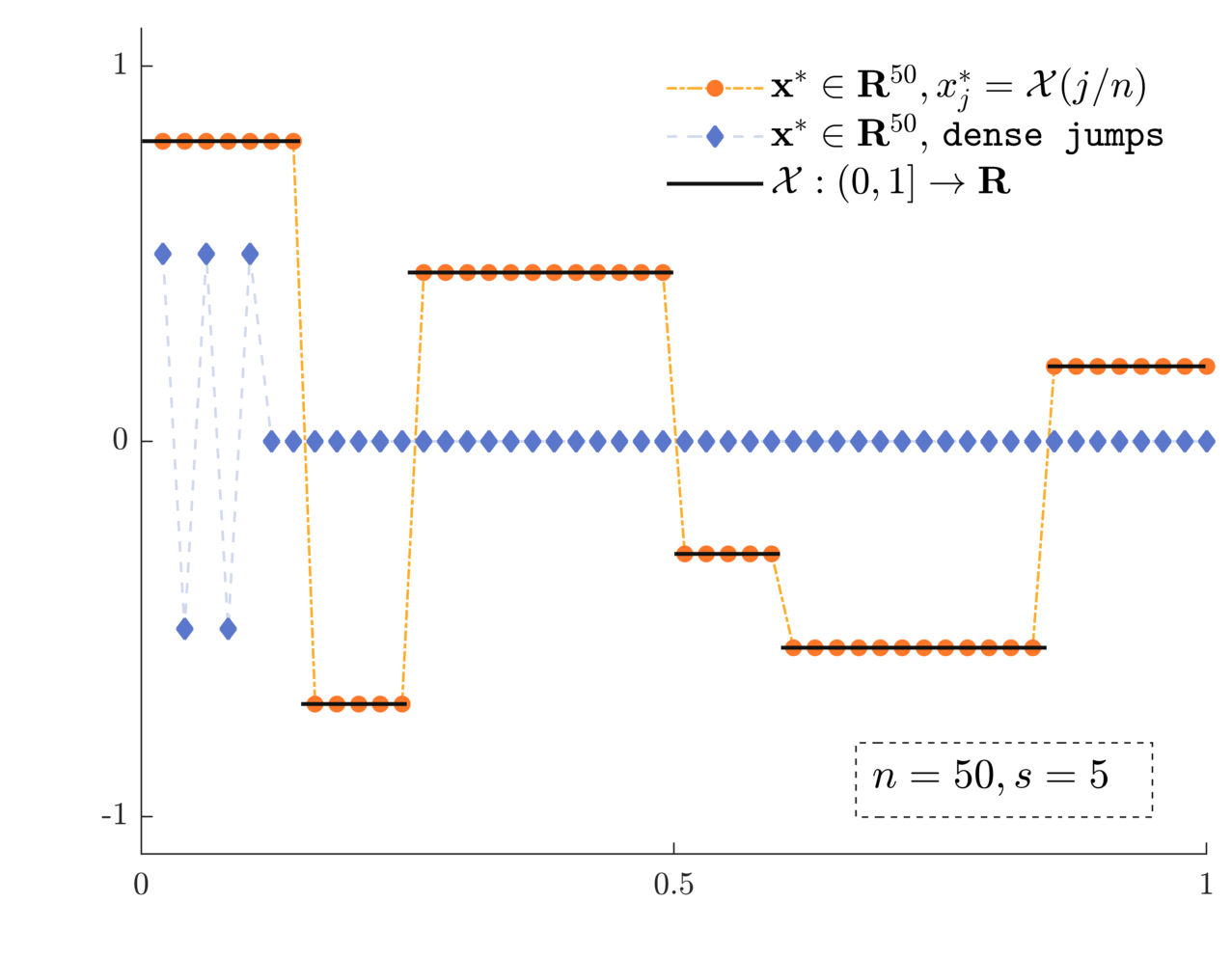}
		\caption{}
		\label{fig:intro:n50}
	\end{subfigure}
	\vspace{1em}
	\begin{subfigure}[t]{0.45\linewidth}
		\centering
		\includegraphics[width=\linewidth]{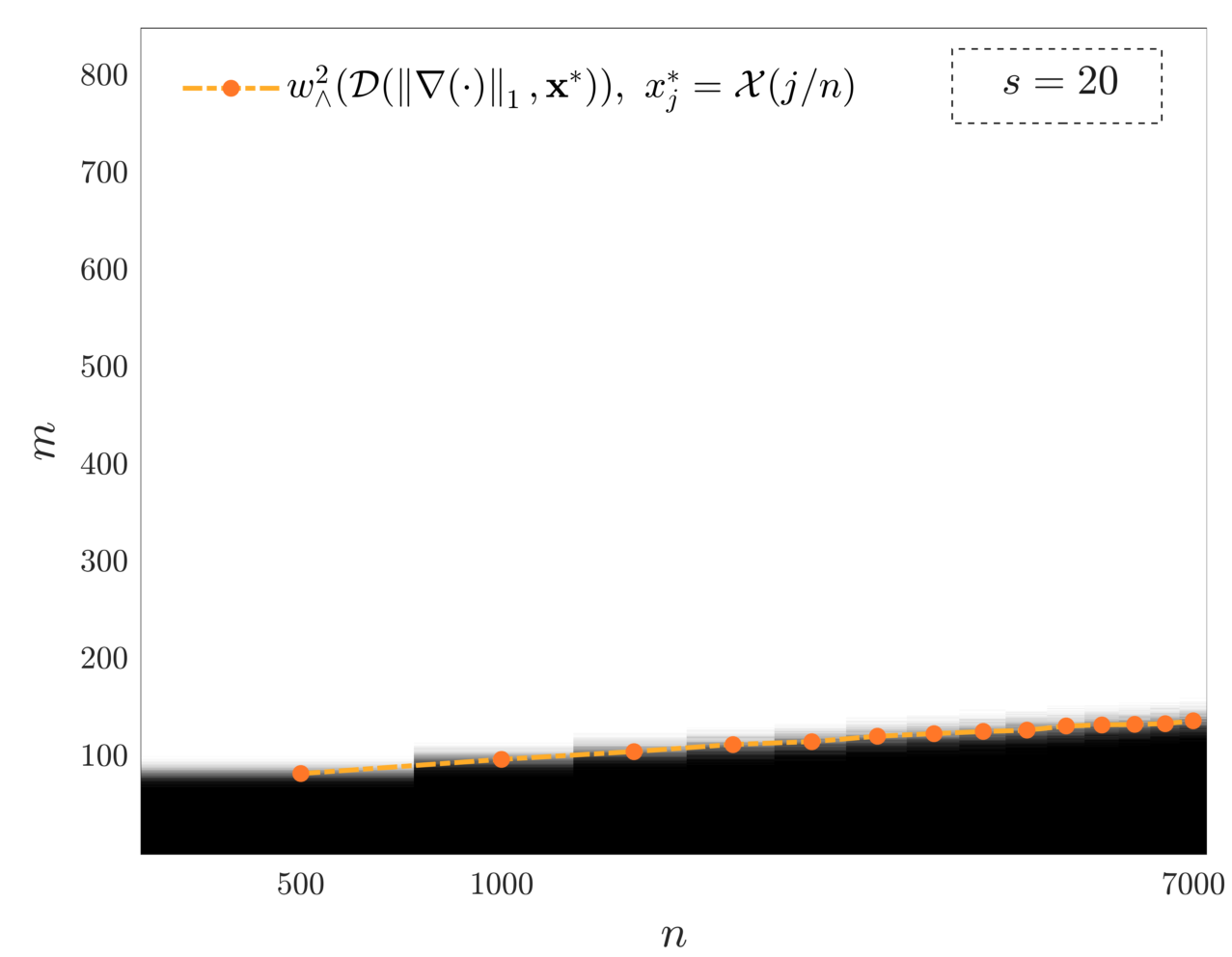}
		\caption{}
		\label{fig:intro:max}
	\end{subfigure}%
	\qquad
	\begin{subfigure}[t]{0.45\linewidth}
		\centering
		\includegraphics[width=\linewidth]{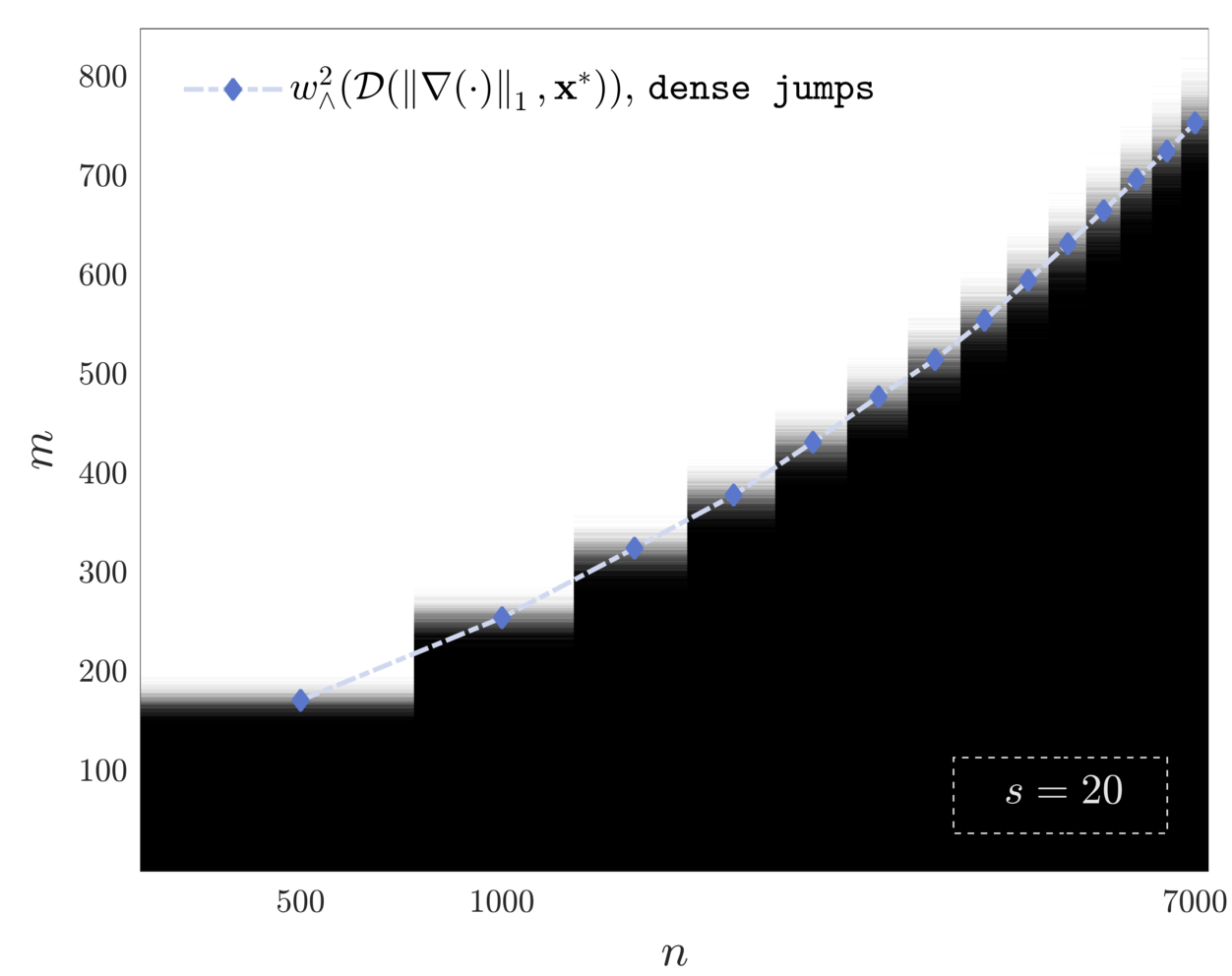}
		\caption{}
		\label{fig:intro:dense}
	\end{subfigure}%
	\caption{\textbf{Numerical simulation.} Subfigure~\subref{fig:intro:n30} and~\subref{fig:intro:n50} show schematic examples of the signal classes that are considered in this experiment at different resolution levels. The orange signal (with circle symbols) is defined as discretization of the piecewise constant function $\contgrtr \colon \intvopcl{0}{1} \to \R$ with $\s = 5$ jump discontinuities that is plotted in black. The blue plot (with diamond symbols) shows a so-called \emph{dense-jump signal}, which does not match the intuitive notion of a $5$-gradient-sparse signal; note that the spatial location of the jumps is chosen adaptively to the resolution level here, which does not correspond to a discretization of a piecewise constant function. For each signal class we have created phase transition plots: Subfigure~\subref{fig:intro:max} and~\subref{fig:intro:dense} display the empirical probability of successful recovery via TV minimization~\eqref{eq:intro:tv-1} for different pairs of ambient dimension $n$ and number of measurements $m$; note the horizontal axis uses a logarithmic scale. The corresponding grey tones reflect the observed probability of success, reaching from certain failure (black) to certain success (white). Additionally, we have estimated the conic Gaussian mean width of $\lnorm{\TV(\cdot)}[1]$ at $\grtr$ (denoted by $\effdim[\conic]{\descset{\lnorm{\TV(\cdot)}[1], \grtr}}$), which is known to precisely capture the phase transition (cf.~\cite{amelunxen2014edge}). The result of Subfigure~\subref{fig:intro:dense} confirms that the class of dense-jump signals suffers from the $\sqrt{\s n}$-bottleneck as predicted by~\cite{cai_guarantees_2015}. On the other hand, Subfigure~\subref{fig:intro:max} reveals that this bottleneck can be broken for discretized signals, as predicted by Theorem~\ref{thm:results:exact}.}
	\label{fig:intro}
\end{figure}

The significance of Theorem~\ref{thm:results:exact} depends on the size of the separation constant $\SC$. In particular, we obtain the near-optimal rate of $m \gtrsim \s \cdot \PolyLog (n)$ if $\SC$ can be chosen independently of~$n$ and $s$.
A typical example of such a situation is the discretization of a suitable piecewise constant function $\contgrtr \colon \intvopcl{0}{1} \to \R$. Indeed, based on Theorem~\ref{thm:results:exact}, \cite[Cor.~2.6]{Genz2020} shows that $m \gtrsim \s \cdot \log^2(n)$ measurements are sufficient for exact recovery when $\contgrtr$ is finely enough discretized; see Figure~\ref{fig:intro} for a visualization of this result.

\vspace{-1.5\baselineskip}
\section{Discussion and Outlook}
We have shown that the $\sqrt{sn}$-bottleneck for 1D TV recovery from Gaussian measurement can be broken for signals with well separated jump discontinuities. The results of Table~\ref{tab:intro} suggest that TV minimization in one spatial dimension plays a special role in this regard. However, we argue that such a phenomenon can also be observed in higher spatial dimensions. In fact, we conjecture that the common rate of $m\gtrsim s \cdot \PolyLog(n,s)$ only reflects worst-case scenarios, while it can be significantly improved for natural signal classes, such as piecewise constant functions with sufficiently smooth boundaries.

% %% You can make the bibliography smaller
% \begin{thebibliography}{10}
% \bibitem{dummyref}
% A. Gebric and H. Armonic, 
% \newblock ``An introduction to humorous mathematics'',
% \newblock The Journal of Very Bad Jokes, {\bf 6}(4):123--134, 2006
% 
% \bibitem{lorem}
% C. Adams
% \newblock ``What does the filler text "lorem ipsum"
% mean?'', The Straight Dope, 2001
% 
% \end{thebibliography}
%%% Biblography

\newpage
\renewcommand*{\bibfont}{\smaller}
% \nocite{*} % auch nichtzitierte Werke im Verzeichnis
\begin{refcontext}[sorting=nyt]
	\printbibliography
\end{refcontext}

\end{document}